\documentclass[a4paper,fleqn]{cas-dc}

\usepackage[numbers,sort&compress]{natbib}
\usepackage{capt-of}        
\usepackage{chemformula}    

\begin{document}
\let\WriteBookmarks\relax

\setcounter{topnumber}{3}
\setcounter{dbltopnumber}{3}
\setcounter{totalnumber}{5}
\renewcommand{\topfraction}{0.95}
\renewcommand{\dbltopfraction}{0.95}
\renewcommand{\textfraction}{0.07}
\renewcommand{\floatpagefraction}{0.7}
\renewcommand{\dblfloatpagefraction}{0.7}

\shortauthors{J. Lee et~al.}

\title[mode = title]{How anion composition governs structural frameworks and Li-ion transport in oxychloride solid-state electrolytes}

\author[1]{Jiho Lee}
\author[1]{Seungwoo Hwang}
\author[1,2]{Seungwu Han}
\author[3,4]{Jung-Hoon Lee}

\author[5]{Jisu Jung}
\cormark[1]
\ead{jsjung@pknu.ac.kr}

\author[6]{Youngho Kang}
\cormark[1]
\ead{youngho84@inu.ac.kr}

\author[3,7]{Sungwoo Kang}
\cormark[1]
\ead{sung.w.kang@kist.re.kr}

\affiliation[1]{organization={Department of Materials Science and Engineering,
    Seoul National University},
    city={Seoul}, postcode={08826}, country={Korea}}
\affiliation[2]{organization={Korea Institute for Advanced Study},
    city={Seoul}, postcode={02455}, country={Korea}}
\affiliation[3]{organization={Computational Science Research Center, 
    Korea Institute of Science and Technology (KIST)},
    city={Seoul}, postcode={02792}, country={Korea}}
\affiliation[4]{organization={KU-KIST Graduate School of Converging Science and Technology,
    Korea University},
    city={Seoul}, postcode={02792}, country={Korea}}
\affiliation[5]{organization={Department of Materials Science and Engineering,
    Pukyong National University},
    city={Busan}, postcode={48513}, country={Korea}}
\affiliation[6]{organization={Department of Materials Science and Engineering,
    Incheon National University},
    city={Incheon}, postcode={22012}, country={Korea}}
\affiliation[7]{organization={Division of Nanoscience and Technology, KIST School,
    University of Science and Technology (UST)},
    city={Seoul}, postcode={02792}, country={Korea}}

\cortext[cor1]{Corresponding author}

\begin{abstract}
Mixed-anion lithium oxychlorides exhibit high Li conductivity, but the relationship between oxygen content, structural framework, and Li-ion transport remains unclear. Here, crystal structure prediction and molecular dynamics simulations with machine-learning interatomic potentials are used to examine \ch{Li_{$y$}$M$O_{$x$}Cl_{6-2$x$}} for $M$ = Nb and Ta (Group 5) and $M$ = Ti, Zr, and Hf (Group 4). Increasing oxygen content changes isolated chloride-rich polyhedra into one-dimensional chains and subsequently more highly connected two- and three-dimensional frameworks. In the Group 5 series, fast Li diffusion occurs over $x \approx 0.67$--1.5, where the Li sites form a three-dimensional percolating network. Potential of mean force maps and nudged elastic band calculations reveal that fast Li-ion transport requires a low-dimensional framework supporting long-range migration through undercoordinated Li sites with low barriers, together with similar energies across Li sites. Finally, we perform fine-grid compositional search, which identifies \ch{LiNbO_{0.67}Cl_{4.67}}, \ch{LiTaO_{0.75}Cl_{4.5}}, \ch{LiNbO_{1.5}Cl3}, and \ch{LiTaO_{1.5}Cl3} as stable candidates with predicted room-temperature self-diffusivities of up to $\sim 10^{-6}$~cm$^2$/s. These findings establish how oxygen content controls framework and Li-ion transport, guiding the design of lithium oxyhalide electrolytes.
\end{abstract}


\maketitle

\section{Introduction}

All-solid-state batteries can improve safety and energy density relative to conventional Li-ion batteries. This requires solid-state electrolytes (SSEs) that combine fast Li-ion transport with chemical and electrochemical stability~\cite{SSE_review1}. Established SSE material families each satisfy only part of this requirement. Sulfides reach superionic conductivities of $10$--$25$~mS/cm but are moisture sensitive~\cite{SSE_review3,SSE_review4,sulfide_25}. Oxides are chemically robust but often exhibit lower ionic conductivity and limited processability~\cite{SSE_review4,oxide_review}. Halide electrolytes offer high-voltage stability but generally exhibit lower ionic conductivity than sulfides and often rely on expensive metals such as In and Y~\cite{halide,LNOC-LTOC}. This trade-off between different material groups has motivated the search for mixed-anion chemistry that enable transport mechanisms unavailable in conventional single-anion SSEs~\cite{SSE_review2,LTOC-Science}.

Lithium oxyhalides have recently emerged as a promising material group. By incorporating oxygen into a chloride framework, Tanaka \textit{et al.} reported crystalline \ch{LiNbOCl4} and \ch{LiTaOCl4} with room-temperature conductivities of 10.4 and 12.4~mS/cm, respectively~\cite{LNOC-LTOC}. Subsequent studies expanded this family through isovalent metal substitution and aliovalent O/Cl exchange with concurrent Li-content tuning within a common structural framework~\cite{Matter,Coherence}: low-cost Zr-based oxychlorides with conductivities above 1~mS/cm~\cite{Zr-npj}, amorphous oxyhalides in which disorder enhances transport~\cite{Zr-jacs,amorphous-aem}, and analogous Na-based mixed-anion electrolytes~\cite{NNOC}. Most notably, Zhao \textit{et al.} reported \ch{Li3Ta3O4Cl10} with a conductivity of 13.7~mS/cm at $25^\circ$C and stability up to 4.9~V~\cite{LTOC-Science}. Its high conductivity was attributed to continuous tetrahedral--tetrahedral (tet--tet) Li pathways that bypass the high-barrier octahedral sites commonly found in oxide and halide frameworks.

The different oxidation states of O and Cl give rise to two compositional design strategies for tuning the O/Cl ratio. The first preserves the parent framework by replacing a fraction of Cl with O and adjusting the Li content, as in \ch{Li_{1-$x$}NbO_{1-$x$}Cl_{4+$x$}}~\cite{Coherence} and \ch{Li_{2+$x$}ZrCl_{6-$x$}O_{$x$}}~\cite{Zr-npj,zr-natcomm}. The second replaces two Cl ions with one O ion while keeping $2n_{\mathrm{O}}+n_{\mathrm{Cl}}$ constant, where $n_{\mathrm{O}}$ and $n_{\mathrm{Cl}}$ are the numbers of O and Cl atoms per formula unit, yielding the general composition \ch{Li_{$_y$}$M$Cl_{6-2$x$}O_{$x$}} ($M$ = cation). This changes the cation-to-anion ratio, requiring the cation--anion framework to reorganize and introducing additional structural degrees of freedom. The distinct oxychloride frameworks adopted by \ch{Li$M$OCl4} ($M$ = Nb, Ta, $x = 1$) and \ch{Li3Ta3O4Cl10} ($x = 1.33$) illustrate this structural flexibility~\cite{LNOC-LTOC,LTOC-Science}. This strategy thus offers opportunities to discover new oxychloride frameworks that support fast Li-ion transport.

Experimental reports in this framework-varying materials space are likewise limited to a small number of compositions~\cite{LNOC-LTOC,LTOC-Science}, so the structural evolution with oxygen content and its effect on Li-ion transport remain largely unexplored. Previous computational studies have effectively screened chemical substitutions within the established \ch{Li$M$OCl4}-derived framework~\cite{Matter}, but have not addressed the potential change in framework. Addressing these questions requires structural and transport information across a broad compositional and configurational space that is impractical to obtain in advance.

Theoretical crystal structure prediction (CSP) offers a route to this problem by searching low-energy structures for a given composition, enabling the discovery of previously unreported materials~\cite{airss,uspex,pso}. This approach, however, requires large amounts of density functional theory (DFT) calculations for structural optimizations, which results in high computational costs. Therefore, recent approaches are developed to replace DFT calculations with machine-learning interatomic potentials (MLIPs)~\cite{bpnn,mtp,gap}, which are machine-learning surrogate models that predict energies and forces after training on DFT datasets. Owing to this near-DFT accuracy at a small fraction of the cost, MLIPs have been widely adopted in CSP to relax and rank the large number of candidate structures generated during the search~\cite{csp-mlip1,csp-mlip2,csp-mlip3}. MLIPs can also accelerate diffusivity calculations, which requires long molecular dynamics (MD) simulations. Pretrained universal MLIPs, trained on large databases spanning much of the periodic table, enable calculations of diverse chemical and structural environments without retraining a separate potential for each composition~\cite{M3GNet,CHGNet,MACE_MP_0,sevennet,eSen,Grace,omat,gnn_Kang}. The use of pretrained models reduces the time required to generate the training set and validate the trained model. Indeed, the state-of-the-art universal MLIPs are known to predict the diffusivities of SSEs well~\cite{topic,sevennet-nano,LPS-benchmark}.

Here, we combine MLIP-accelerated CSP and MD simulations to explore the halide-to-oxide compositional range of \ch{Li_{$y$}$M$O_{$x$}Cl_{6-2$x$}} for Group 5 metals ($M$ = Nb, Ta, and y = 1) and Group 4 metals ($M$ = Ti, Zr, Hf, and y = 2) from $x = 0$ to 3. MD simulations of the ten lowest-energy structures at each composition reveal the structural and transport trends across this range. For Nb and Ta, increasing oxygen content connects the metal--anion framework and changes the local coordination environment of Li, thereby governing Li-ion transport across the compositional range. In contrast, Ti, Zr, and Hf exhibit distinct structural trends at intermediate oxygen contents, leading to a substantially narrower range of stable compositions. Potential of mean force (PMF) \cite{pmf} maps and climbing-image nudged elastic band (CI-NEB) \cite{CI-neb} calculations show that the $x \approx 0.67$--1.5 Group 5 compositions contain low-coordination Li sites connected by a shallow, three-dimensionally percolating migration landscape. These results establish a structural design rule for lithium oxyhalides and identify intermediate-oxygen Nb and Ta compositions as candidates for fast solid-state electrolytes.

\section{Results and discussion}

\subsection{Structure prediction workflow and validations}
\label{sec:prediction}

We identify the structures of \ch{Li_{$y$}$M$O_{$x$}Cl_{6-2$x$}} using CSP. We first scan Group 5 metals Nb and Ta ($y = 1$) and Group 4 metals Ti, Zr, and Hf ($y = 2$) from $x = 0$ to $x = 3$, using intervals of $\Delta x = 0.5$ (section~\ref{sec:oxygen}). Then, we perform screening on a finer $\Delta x$ grid around promising candidates in section~\ref{sec:fine}. As only three oxychlorides in these series have been reported experimentally, namely \ch{LiNbOCl4} and \ch{LiTaOCl4} ($x = 1$) and \ch{Li3Ta3O4Cl10} ($x = 1.33$)~\cite{LNOC-LTOC,LTOC-Science}, the structures across most of the composition range remain unknown. For each composition, candidate structures are generated using SPINNER, a CSP method that combines random structure generation and evolutionary operations with local optimization and energy evaluation performed with MLIPs~\cite{spinner,csp_mqa,CSP_benchmark}. Without using any structural prototype, SPINNER identifies the experimental or more stable phases for about 80\% of 60 ternary compositions in blind tests, and the same approach has been used to explore 181 previously uncharted ternary oxide systems~\cite{spinner,spinner_oxide}. Candidate crystal structures are generated through CSP using bespoke MLIPs trained with disordered structures, then reoptimized and ranked by energy using the pretrained universal potential SevenNet-Omni~\cite{sevennet-omni}. Li-ion diffusivities are then calculated from mean-square displacements obtained in MD simulations using SevenNet-Omni. The performance of SevenNet-Omni has been benchmarked across diverse inorganic materials, including for Li-ion transport in solid-state electrolytes (SSEs)~\cite{sevennet-nano,Omni_benchmark}. For accurate description of the lattice parameters, we include D3 van der Waals (vdW) corrections~\cite{D3, dimensionality,LNOC-d3}. The computational methods and parameters used for SPINNER and Li-ion transport calculations are detailed in the Methods section.

As a validation, we evaluate SPINNER's performance for chloride, oxide, and oxychloride compositions by comparing the generated structures with experimentally reported structures (Fig.~S1). We find that SPINNER successfully identifies the experimental structural frameworks, except for \ch{Li3Ta3O4Cl10} (Table~S1). For the three oxide compositions (\ch{Li2ZrO3}, \ch{LiNbO3}, \ch{LiTaO3}), SPINNER generates the corresponding experimental crystal structures, including the Li positions. For \ch{Li2ZrCl6} and \ch{LiNbOCl4}, which exhibit Li-site disorder due to partial Li occupancy, SPINNER successfully recovers the corresponding non-Li frameworks.  \ch{Li3Ta3O4Cl10}, which has a large primitive cell containing 80 atoms, is the only tested composition for which SPINNER does not identify the experimental structure. Nevertheless, the lowest-energy SPINNER candidate is only 7~meV/atom higher in energy than the experimental structure and exhibits similar structural characteristics (Fig.~S2). Both structures contain one-dimensional metal--anion frameworks and show similar radial distribution functions, bond-angle distributions, and coordination numbers (Figs.~S2a--d). The candidate and experimental structures also show Li diffusivities and transport characteristics that are similar to each other (Figs.~S2e and f), consistent with their similar metal-polyhedron motifs and framework connectivity. 

It is important to note that our goal is not to predict the exact experimentally observed structures that are synthesizable. Rather, we aim to predict material properties and use them to identify promising compositions. In some cases, SPINNER identifies structures with lower MLIP energies than the experimental structures (Fig.~S1). This demonstrates that SPINNER effectively optimizes structures with respect to the given MLIP energy model, but it does not ensure that these lower-energy structures are experimentally accessible. Predicting which structures form experimentally remains a separate challenge because DFT errors, the disorder effects from partial occupancy of Li sites, and finite-temperature contributions can alter the relative energy ordering of candidate structures. Moreover, experimentally observed structures depend on the energetics and kinetics of the synthesis pathway.  Nevertheless, as shown above, low-energy candidates identified by SPINNER share similar structural features with experimentally realized systems, leading to similar Li-transport characteristics. These predictions can therefore guide subsequent experiments aimed at optimizing compositions.

We next evaluate the accuracy of SevenNet-Omni with D3 corrections by comparing with DFT calculations (PBE functional \cite{pbe} with D3 corrections) for structures spanning all five metal elements and oxygen contents (shown in Fig.~S3). SevenNet-Omni+D3 reproduces the DFT energies and forces with RMSEs below 6~meV/atom and 0.07~eV/\AA{}, respectively. In addition, we check the accuracy of SevenNet-Omni+D3 for Li diffusivity calculations. We first observe that the diffusivity values are converged with respect to supercell size when all lattice parameters exceed 10~\AA{} (Fig.~S4a). In addition, the Arrhenius relation calculated for \ch{LiNbOCl4} with SevenNet-Omni+D3 agrees with two independent DFT-MD results (Fig.~S4b)~\cite{Matter,LNOC-JACS}. 

\subsection{Trends in structure and Li diffusivity}
\label{sec:oxygen}

\begin{figure*}[!t]
\centering
\includegraphics[width=0.85\textwidth]{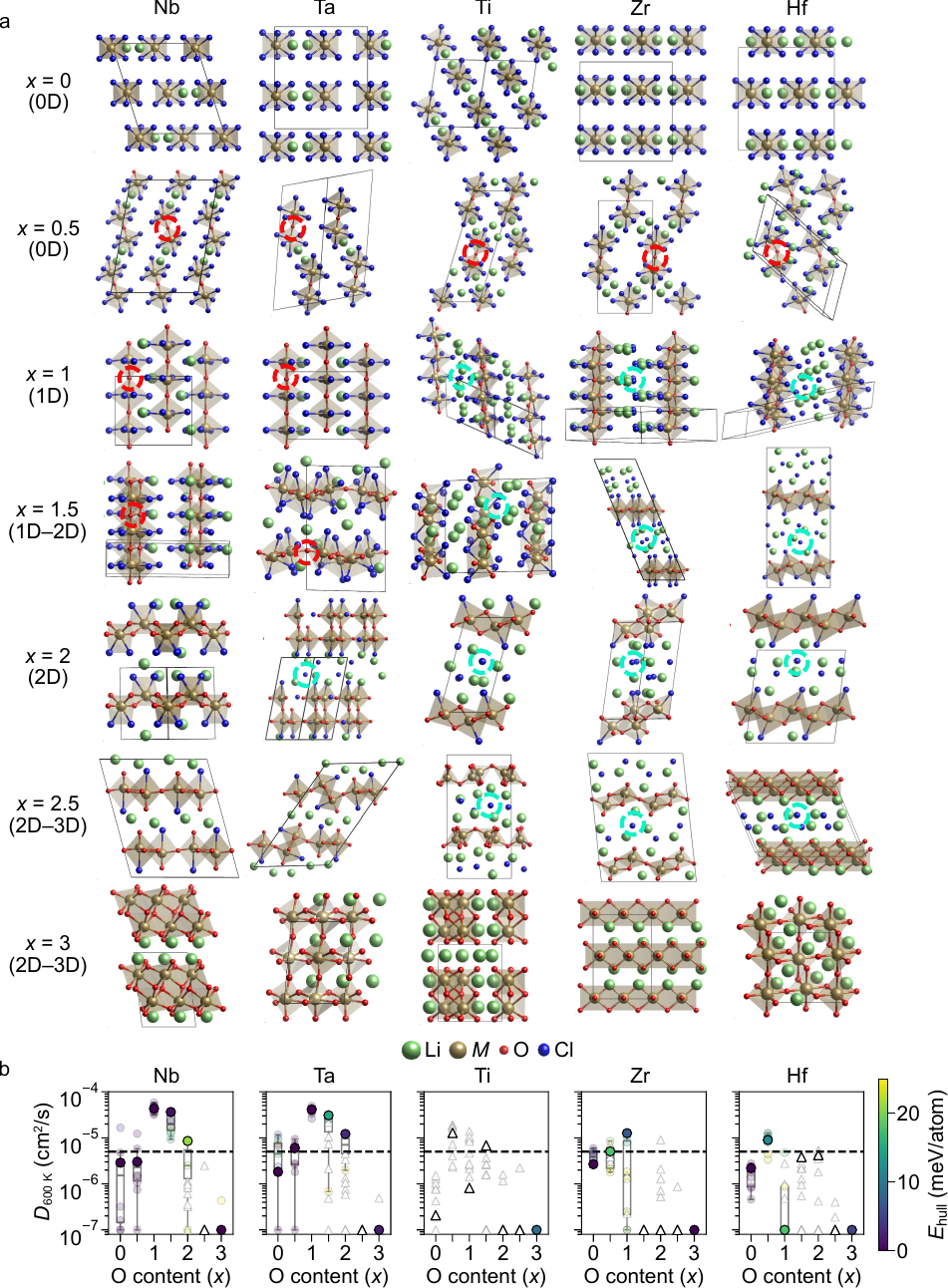}
\caption{\textbf{Structures and Li-ion transport in \ch{Li_{$y$}$M$O_{$x$}Cl_{6-2$x$}}.} (a) Lowest-energy structures of \ch{Li$M$O_{$x$}Cl_{6-2$x$}} ($M$ = Nb, Ta) and \ch{Li2$M$O_{$x$}Cl_{6-2$x$}} ($M$ = Ti, Zr, Hf) for $x = 0$--3. Red circles indicate corner-sharing $M$--O--$M$ linkages between metal-centered polyhedra, and blue circles indicate Cl sites with no metal coordination. (b) 600~K Li diffusivity for the ten lowest-energy structures at each composition of the Nb, Ta, Ti, Zr and Hf series. Symbol color indicates $E_{\mathrm{hull}}$. Open triangles identify structures above the 25~meV/atom stability window, and boxes show the range of hull-accessible structures. The dashed horizontal line marks the diffusivity threshold of $D_{600\,\mathrm{K}} = 5\times10^{-6}$~cm$^2$/s.}
\label{fig:overview}
\end{figure*}

Fig.~\ref{fig:overview}a shows the lowest-energy structures of the Group 5 (Nb and Ta) and the Group 4 (Ti, Zr, and Hf) compounds predicted from SPINNER across the halide-to-oxide compositional range. The most noticeable change is the evolution of the dimension of the structural framework from 0D ($x = 0$) into a 2D/3D network ($x = 3$). At $x = 0$, all five metals form isolated \ch{[$M$Cl6]} octahedra within a chloride ionic packing, with no links between neighboring metal-centered polyhedra. At $x = 0.5$, the structures retain 0D units consisting of coupled polyhedra bridged by an O atom. The statistics in Fig.~S5 show that O primarily acts as a bridging anion coordinated to multiple $M$ cations, whereas Cl remains terminal with respect to the $M$-centered polyhedral framework. Accordingly, the frameworks change into 1D chains at $x = 1.0$, and gradually transition to 2D layers and 3D networks as the O content increases.

During the structural evolution with increasing $x$, the connectivity between the $M$ and O atoms changes, with corner-sharing motifs predominating at lower $x$ and edge-sharing motifs emerging alongside them as $x$ increases. The evolution, however, differs between the two metal groups: Group 5 compounds retain exclusively corner-sharing frameworks at $x = 1$, whereas Group 4 compounds already begin to incorporate edge-sharing motifs at this composition, forming either corner-sharing or mixed frameworks. For example, the minimum-energy structure of \ch{Li2ZrOCl4} adopts edge-sharing connectivity, while competing corner-sharing phases lie more than 67 ~meV/atom above it (Fig.~S6). In contrast, for \ch{LiNbOCl4}, the minimum-energy corner-sharing structure is more stable than the edge-sharing structure by 22~meV/atom. The preference for corner-sharing motifs in Group 5 can be attributed to stronger electrostatic repulsion between its more highly charged cations than between Group 4 cations, which makes edge-sharing motifs less favorable because of their shorter cation--cation distances.

The coordination environment of Cl also evolves with increasing $x$ (see blue circles in Fig.~\ref{fig:overview}a). Direct coordination of Cl to $M$ cations decreases, whereas the number of Cl ions surrounded exclusively by Li ions through ionic interactions increases. (The distinct bonding environments are supported by the Bader charge analysis~\cite{bader} shown in Fig.~S7.) Notably, Group 5 compounds retain more direct $M-$Cl coordination at higher $x$ than Group 4 compounds. This may result from stronger electrostatic attraction between Group 5 metals and Cl, as well as the more open $M-$O framework at corner-sharing motifs than edge-sharing structures. The open framework reduces anion--anion repulsion around Cl and thereby helps preserve direct $M-$Cl coordination. Relatively lower Li contents in Group 5 compounds can further contribute to this trend by reducing the number of Li ions available to surround and stabilize Cl.

Fig.~\ref{fig:overview}b shows the hull energies ($E_{\mathrm{hull}}$) and Li diffusivities of the ten lowest-energy candidates for each composition. In Li-containing electrolytes, Li ions may partially occupy multiple crystallographic sites with similar energies, giving rise to site disorder through configurational entropy. To account for this effect, we use an energy buffer of 25~meV/atom, approximately equal to $k_{\mathrm{B}}T$ at room temperature, to assess the thermodynamic stability of the candidate compounds. We find that the Group 4 series contains few compounds in the stability window within 25~meV/atom. Zr and Hf have accessible structures only at $x \le 1$ and $x = 3$, and Ti has essentially none except for $x = 3$. In contrast, Nb and Ta contain stable compounds across much wider composition range (see also Fig.~S8 for detailed statistics).

For Li diffusivity (Fig.~\ref{fig:overview}b), we calculate the scalar diffusivity, $D$, from the total mean-square displacement (MSD) at 600~K ($D_{600\,\mathrm{K}}$), a temperature chosen to capture Li-ion transport within the accessible MD timescale. The 300~K diffusivities of selected promising candidates are obtained separately. The reported values are averaged over three independent simulations. All metal groups exhibit a volcano-shaped trend, with the highest diffusivities occurring for $x = 0.5$--1.5, where the structures are one-dimensional (1D). These peak compositions have $D_{600\,\mathrm{K}} > 5\times10^{-6}$~cm$^2$/s, corresponding to estimated values of $D_{300\,\mathrm{K}} > 1.5\times10^{-8}$~cm$^2$/s and $\sigma_{300\,\mathrm{K}} > 0.5$~mS/cm when an activation energy ($E_{\mathrm{a}}$) of 0.3~eV, typical of SSEs, is assumed~\cite{SSE_Ea}. In contrast, 0D structures at $x = 0$, as well as oxygen-rich structures at $x \geq 2$, exhibit much lower diffusivities, indicating nearly immobile Li ions. At the diffusivity peak, Group 5 compounds exhibit higher Li-ion diffusivities than Group 4 compounds.

\subsection{Structural factors affecting diffusivity}
\label{sec:mechanism}

\begin{figure*}[!t]
\centering
\includegraphics[width=0.8\textwidth]{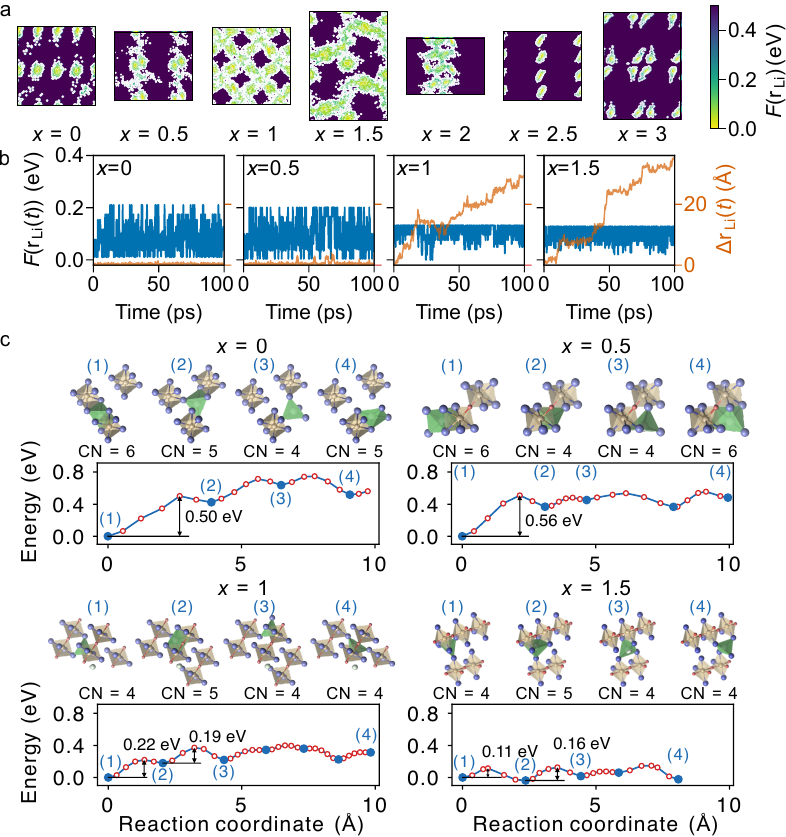}
\caption{\textbf{Li free-energy landscape and migration barriers in \ch{LiNbO_{$x$}Cl_{6-2$x$}}.}
(a) Potential of mean force $F(\mathbf{r}_{\mathrm{Li}})$ obtained from the Li density sampled during 600~K MD of the lowest-energy structure at each $x$. The maps are shown as projections, with the most populated Li site referenced to $F = 0$. The color scale is truncated at 0.5~eV and the dark color regions mark the parts of the cell that Li does not reach during the simulation. (b) $F$ evaluated along the trajectory of a single Li ion for $x = 0$--1.5. The trajectory is colored
by the net displacement $\Delta \mathbf{r}_{\mathrm{Li}}(t)$ from the initial position on the right axis. (c) CI-NEB energy profiles for Li migration through consecutive sites. Filled symbols denote the initial and final states, and open symbols denote the intermediate NEB images. The structures above each profile show the local environment of the migrating Li at the labeled sites, with its coordination number indicated.}
\label{fig:mechanism}
\end{figure*}

To identify the structural factors underlying the high Li-ion diffusivities of Group 5 compounds near the peak compositions, we analyze Li-ion diffusion pathways and their associated energetics as a function of $x$. As representative low-energy structures for each composition, we select the lowest-energy Nb- and Ta-containing candidates and calculate the potential of mean force (PMF)~\cite{pmf}, $F(\mathbf{r}_{\mathrm{Li}})$, from their 600~K MD trajectories. The PMF converts the time-averaged Li-ion density into a free-energy landscape through the inverse Boltzmann relation (see Methods section for details), with its minimum, corresponding to the maximum Li-ion density, set to zero. Fig.~\ref{fig:mechanism}a and Fig.~S9 show the PMF maps for the Nb- and Ta-containing compound families, respectively, which exhibit similar trends. The maps are projected onto selected crystallographic planes to visualize Li-ion migration pathways. At $x = 1$ and 1.5, regions with PMF values below 0.2~eV form continuous pathways across the crystal, indicating that Li ions can effectively percolate along energetically favorable routes. At the other $x$ values, these low-PMF regions are less connected, hindering long-range Li-ion migration. We observe that, at $x = 0$ and 0.5, the Li density remains localized around isolated six-coordinate, Cl-rich sites between the \ch{NbCl6} and \ch{Nb2OCl10} units. At $x=2$ and 2.5, the increasingly compact, predominantly two-dimensional $M$--anion frameworks restrict Li-ion migration within the frameworks. Although Li ions can migrate to some extent in the in-plane regions between adjacent frameworks, transport across the frameworks in the perpendicular direction is hindered, disrupting three-dimensional long-range transport and confining Li ions to localized regions. At the highest oxygen contents ($x=3$), where the framework becomes nearly three-dimensional, the compact structure further restricts Li-ion migration. 


We further examine the temporal evolution of $F$ and Li positions for compositions with $x < 2$. These compositions are more stable than those with $x \geq 2$ in most cases, whereas the highest Li-ion diffusivities occur near $x = 1$ and 1.5. In Fig.~\ref{fig:mechanism}b, the PMF evolution is obtained by tracking an individual Li ion on the PMF landscape derived from the density distribution of all Li ions in Fig.~\ref{fig:mechanism}a; $\Delta \mathbf{r}_{\rm Li}$ denotes the displacement of that ion. At $x \leq 0.5$, the PMF value along the tracked Li-ion trajectory oscillates between 0 and approximately 0.2~eV, while the Li ion moves by approximately 1~\AA{} or less over the 100~ps trajectory. This indicates that the Li ion remains localized at a Cl-rich site with a coordination number (CN) of 5--6. The small deviations of the PMF from zero arise from the finite spatial resolution of the grid-based Li-density analysis and do not reflect physical energy differences between sites.  At $x = 1$ and 1.5, Li ions hop continuously between sites with coordination numbers of 4--5, and their trajectories extend over tens of \AA{}. In contrast to $x = 0$ and 0.5, the free-energy variation along these migration pathways remains below 0.15~eV, indicating that Li ions can percolate through a relatively low-energy landscape.

To directly assess the energetics along Li-ion migration pathways, we calculate activation barriers using reaction sampling and climbing-image nudged elastic band (CI-NEB) calculations~\cite{CI-neb}. Due to the concerted Li-ion migration, it is difficult to obtain candidate event pathways of Li diffusion in SSEs. Therefore, we use the ARTn automated saddle-point-search method~\cite{sherpa} to identify unbiased migration pathways involving correlated motion of multiple Li ions. We employ SevenNet-Nano~\cite{sevennet-nano}, a lightweight student model of SevenNet-Omni, to reduce the computational cost of this search. The resulting pathways are then refined using CI-NEB calculations with SevenNet-Omni to obtain more accurate activation barriers. Fig.~\ref{fig:mechanism}c shows the CI-NEB calculation results. At $x = 0$ and 0.5, the first hop out of the six-coordinate Li site, which is the lowest-energy site, has a relatively high barrier of 0.50 and 0.56~eV, respectively. This indicates that Li ions are kinetically trapped at these sites, consistent with the negligible Li displacement over the simulation time (Fig.~\ref{fig:mechanism}b). In contrast, at $x = 1$ and 1.5, Li ions occupy four-coordinate sites in the ground states and have lower barriers for leaving their initial positions: 0.22 and 0.11~eV, respectively. Subsequent hops along the long-range migration pathways also have low barriers, resulting in larger Li displacements in the MD simulations (Fig.~\ref{fig:mechanism}b). The calculated barrier for \ch{LiNbOCl4} is consistent with previously reported activation energies of approximately 0.2~eV~\cite{LNOC-LTOC,LNOC-JACS}.

Based on the analyses above, the composition-dependent Li-ion transport in Group 5 compounds can be understood through the interplay between Li-site coordination and framework connectivity. At low $x$ ($x \leq 0.5$), Li ions are stabilized at sixfold-coordinated, Cl-rich sites that correspond to pronounced energy minima. Migration from these sites requires Li ions to pass through lower-coordination configurations, leading to a substantial energy increase. At high $x$ ($x \geq 2$), the increasingly connected and compact $M-$O framework may impose a larger energetic penalty on the local structural rearrangements accompanying Li-ion migration, thereby hindering the formation of continuous low-energy pathways. By contrast, at intermediate $x$ ($x = 1$ and 1.5), the reduced Cl content than $x$ = 0, 0.5 lowers Li--Cl coordination, while the $M$--O framework remains relatively open than $x \geq 2$. Li-ion migration consequently involves only modest changes in Li coordination and limited distortion of the $M-$O framework, resulting in small free-energy variations along the pathways. This balance between Li-site coordination and $M-$O framework connectivity underlies the enhanced Li-ion transport at intermediate compositions.

\subsection{Fine-grid materials screening}
\label{sec:fine}

\begin{figure*}[!t]
\centering
\includegraphics[width=0.85\textwidth]{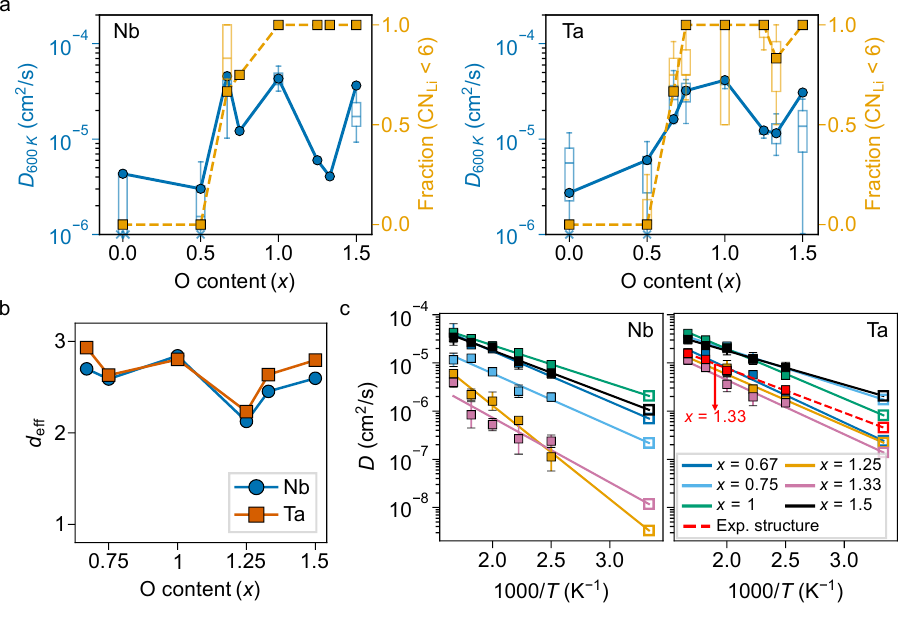}
\caption{\textbf{Fine compositions in the fast-conduction window.}
(a) Li diffusivity at 600~K, $D_{600\,\mathrm{K}}$, and the fraction of Li ions with a coordination number below six for \ch{Li$M$O_{$x$}Cl_{6-2$x$}} ($M=$ Nb, Ta) over $x = 0$--1.5. Blue symbols refer to the left axis and orange symbols refer to the right axis. Boxes show the distribution over the hull-accessible structures of each composition. (b) Effective dimensionality of Li diffusion, $d_{\mathrm{eff}}$, obtained from the eigenvalues of the 600~K diffusion tensor of the same structures, over $x = 0.67$--1.5. (c) Arrhenius plots of the lowest-energy hull-accessible structures at $x = 0.67$--1.5. Filled squares are calculated from 600 to 400~K and error bars show the standard deviation over the independent runs, while the lines are the Arrhenius fits extended to 300~K and the open squares mark the extrapolated $D_{300\,\mathrm{K}}$. Red dashed line denotes the experimentally reported structures calculated with SevenNet-Omni+D3.}
\label{fig:fine}
\end{figure*}

For a more detailed assessment of SSE candidates, we further examine Group 5 compounds at $x = 0.67$, 0.75, 1.25, and 1.33. The lowest-energy structures identified by SPINNER at these compositions are shown in Fig.~S10. They follow the previously discussed trends in the dimensionality of the $M$--O framework and the Li coordination environment as $x$ varies. In particular, the fraction of Li ions with CNs below 6 generally increases with $x$ for compounds with $E_{\rm hull} < 25$~meV, as shown for the Nb and Ta series in Figs.~\ref{fig:fine}a and b, respectively. When more than ten compounds satisfy this hull-energy criterion, we limit the analysis to the ten lowest-energy compounds. At low $x$, the fraction of Li ions with CNs below 6 increases markedly between $x = 0.5$ and 0.67, accompanied by a several-fold increase in the scalar diffusivity at $x = 0.67$. Above $x = 0.67$, the diffusivity varies non-monotonically as the $M$--O framework develops, but it fluctuates without falling below the values observed at $x \leq 0.5$.

To assess whether these compositions support three-dimensional Li-ion transport, which is less sensitive to crystallographic orientation and may facilitate Li-ion transport across randomly oriented grains, we quantify the effective transport dimensionality ($d_{\mathrm{eff}}$) from the diffusivities along three mutually orthogonal principal axes (see Methods section for details). Here, $d_{\mathrm{eff}} = 1$ corresponds to one-dimensional transport, $d_{\mathrm{eff}} = 2$ to isotropic two-dimensional transport, and $d_{\mathrm{eff}} = 3$ to isotropic three-dimensional transport. Fig.~\ref{fig:fine}c shows $d_{\mathrm{eff}}$ for the lowest-energy Nb- and Ta-containing compounds between $x=0.67$ and 1.5. For $x \geq 0.5$, both Group 5 compounds exhibit $d_{\mathrm{eff}} = 2.6$--2.9, except at $x = 1.25$, where the values are slightly lower (2.13 for Nb and 2.23 for Ta). These results suggest that the identified compounds with $x = 0.67$--1.5 largely support Li-ion percolation along all three crystallographic directions. Consistent with this prediction, a previous study reported high Li diffusivity and three-dimensional transport in \ch{LiNbOCl4}, whose lowest-energy structure identified by our CSP analysis matches the experimentally reported structure~\cite{LNOC-JACS}.


We next estimate the room-temperature diffusivities of the candidates. In addition to the 600~K MD simulations, we perform simulations at 550 and 500~K for 200~ps and at 450 and 400~K for 400~ps, using three independent runs at each temperature. Arrhenius fits to the resulting diffusivities are shown in Fig.~\ref{fig:fine}d. As discussed above, the structure examined for \ch{LiNbOCl4} ($x = 1$) matches the experimental structure. For \ch{Li3Ta3O4Cl10} ($x = 1.33$), we additionally evaluate the experimentally reported structure alongside the computationally identified structure for comparison. Between 400 and 600~K, the diffusivities show no evidence of the non-Arrhenius behavior reported for several SSEs~\cite{dimensionality,LGPS-nonArrh,JH_argyrodite}. Consistently, the 400~K PMF maps for the $x = 1$ and 1.5 structures in both the Nb and Ta series retain percolating low-$F$ networks similar to those at 600~K (Fig.~S11), indicating that the diffusion mechanism remains unchanged over the simulated temperature range. This supports extrapolating the higher-temperature diffusivities to 300~K, although a change in the transport mechanism at 300~K cannot be completely ruled out.

\begin{center}
\includegraphics[width=0.8\columnwidth]{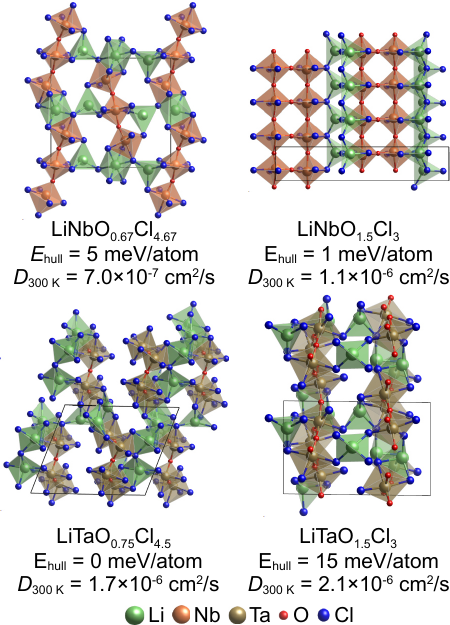}
\captionof{figure}{\textbf{Structures of the proposed candidates.} Lowest-energy structures of \ch{LiNbO_{$x$}Cl_{6-2$x$}} (top row) at $x = 0.67$ and 1.5 and of \ch{LiTaO_{$x$}Cl_{6-2$x$}} (bottom row) at $x = 0.75$ and 1.5. The energy above the convex hull and the extrapolated room-temperature diffusivity are given below each structure.}
\label{fig:final_structures}
\end{center}

For the lowest-energy Nb compounds, the structures at $x = 0.67$, 1, and 1.5 follow Arrhenius behavior, with low activation energies ranging from 0.165 to 0.208~eV (Table~\ref{tab:gs_summary}). Accordingly, they exhibit relatively high diffusivities at 300~K, exceeding $5\times10^{-7}$~cm$^2$/s, comparable to those of well-known Li-ion conductors such as \ch{Li6PS5Cl} (LPSC)~\cite{LPSC}, \ch{Li10GeP2S12} (LGPS)~\cite{LGPS}, and \ch{LiNbOCl4}~\cite{LNOC-JACS}. The structure at $x = 0.75$ shows a similar $E_{\rm a}$ but a lower prefactor, resulting in lower room-temperature diffusivity. In contrast, the structures at $x = 1.25$ and 1.33 exhibit stronger temperature dependence, with $E_{\mathrm{a}} \approx 0.3$--$0.4$~eV, indicating lower diffusivities at room temperature. For the Ta compounds, the activation energy is relatively insensitive to $x$, remaining below $\sim 0.3$~eV, and exhibit higher diffusitivies compared to the Nb compounds. Among the compositions examined, those with $x = 0.75$, 1, and 1.5 exhibit higher room-temperature diffusivities ($8.2\times10^{-7}-2.1\times10^{-6}$~cm$^2$/s) than experimentally synthesized \ch{Li3Ta3O4Cl10} ($x = 1.33$), whose diffusivity is estimated from an Arrhenius fit to values calculated using the same computational framework (Table~\ref{tab:gs_summary}).

\begin{table*}[!t]
  \centering
  \caption{Calculated properties of the selected \ch{Li$M$O_{$x$}Cl_{6-2$x$}} compositions. *The experimentally reported structure is used for \ch{Li3Ta3O4Cl10} ($x = 1.33$)}
  \label{tab:gs_summary}
  \begin{tabular}{llccccc}
    \toprule
    $M$ & $x$ & ESW (V) & $E_{\mathrm{g}}$ (eV) & $E_{\mathrm{a}}$ (eV)
        & $D_{300\,\mathrm{K}}$ (cm$^2$/s) & $\sigma_{300\,\mathrm{K}}$ (mS/cm) \\
    \midrule
    Nb & 0.67 & 2.75--4.05 & 3.33 & 0.208 & $7.0\times10^{-7}$ & 22.6 \\
       & 1.5  & 2.85--4.05 & 3.47 & 0.183 & $1.1\times10^{-6}$ & 45.2 \\
       & 1    & 2.80--4.05 & 3.86 & 0.165 & $1.8\times10^{-6}$ & 70.9 \\
    \midrule
    Ta & 0.75 & 2.25--4.15 & 3.73 & 0.155 & $1.7\times10^{-6}$ & 60.2 \\
       & 1.5  & 2.25--4.15 & 3.81 & 0.140 & $2.1\times10^{-6}$ & 94.3 \\
       & 1    & 2.20--4.20 & 4.29 & 0.205 & $8.2\times10^{-7}$ & 32.2 \\
       & 1.33* & 2.25--4.15 & 3.78 & 0.182 & $4.6\times10^{-7}$ & 19.7 \\
    \bottomrule
  \end{tabular}
\end{table*}

\subsection{Assessment of newly identified candidates}
\label{sec:fine_properties}

Thus far, we have discussed the composition-dependent transport properties of the \ch{Li_{$y$}$M$O_{$x$}Cl_{6-2$x$}} structures identified by SPINNER, and have shown that Group~5 compounds can be stable as oxyhalides over an appropriate composition range, primarily at $x \leq 2$, while exhibiting high Li-ion diffusivities. We next assess other properties relevant to SSE applications, including ionic conductivity, electrochemical stability window (ESW), and band gap ($E_{\mathrm{g}}$), for Nb compounds with $x = 0.67$, 1, and 1.5 and Ta compounds with $x = 0.75$, 1, and 1.5 (Table~\ref{tab:gs_summary} and Fig.~S12). We select these compositions because they are expected to exhibit better transport properties than the other compositions examined. The corresponding lowest-energy structures are shown in Fig.~\ref{fig:final_structures} and are used in the following analyses.

It should be noted that calculated ionic conductivities can vary substantially depending on the computational methods and simulation settings, as illustrated by the range of reported values for several SSEs in Table~S2. Thus, the calculated values are best interpreted as estimates rather than direct quantitative predictions of experimental conductivity. With this caveat, the conductivities predicted for the compounds identified in this work are comparable to calculated values reported for SSEs that exhibit high experimental conductivities of approximately 1--10~mS/cm. This comparison supports the promise of these oxychlorides as fast Li-ion conductors.

Both Group~5 compounds exhibit calculated ESWs wider than 1~V: 2.75--4.05~V for the Nb compound and 2.25--4.15~V for the Ta compound. These windows are substantially wider than the reported ESWs of well-known sulfide SSEs, such as LGPS and LPSC, which are less than 0.5~V~\cite{esw_benchmark}. This comparison suggests that the oxychloride compounds may offer an advantage over these sulfide SSEs in terms of electrochemical stability. In addition, the hybrid-functional calculations predict band gaps greater than 3.3~eV for both compounds, comparable to the values reported for representative SSEs~\cite{LGPS-gap}. These wide band gaps suggest that electronic conduction is suppressed in the Group~5 oxychlorides.

\section*{Conclusions}

In summary, we map the halide-to-oxide compositional space of \ch{Li_{$y$}$M$O_{$x$}Cl_{6-2$x$}} by combining crystal structure prediction with molecular dynamics based on a pretrained MLIP. We show that Li-ion transport exhibits a volcano-shaped dependence on oxygen content, peaking at $x$ = 1--1.5. Our analysis of the migration landscape identifies two structural factors for fast Li-ion transport. The framework must remain low-dimensional so that it does not hinder long-range migration, and Li must occupy undercoordinated sites with low local migration barriers. These conditions explain why the chloride-rich and oxygen-rich compositions conduct poorly, and they are satisfied only by the corner-linked oxychlorides of the Nb and Ta series at $x \approx 0.67$--1.5. Guided by them, we propose \ch{LiNbO_{0.67}Cl_{4.67}}, \ch{LiTaO_{0.75}Cl_{4.5}}, \ch{LiNbO_{1.5}Cl3}, and \ch{LiTaO_{1.5}Cl3} as thermodynamically accessible candidates with fast Li-ion transport beyond the established $x = 1$ oxychlorides, whereas the Group 4 series favors edge-sharing $M$--O units in the same window and conducts poorly. Overall, our work provides a mechanism-based route to designing lithium oxyhalide electrolytes in a compositional space that has remained largely unexplored.

\section{Methods}
\subsection{Crystal structure prediction}
\label{sec:methods_csp}
 
Crystal structures are predicted using SPINNER, which combines random and evolutionary sampling accelerated by a machine-learning interatomic potential without requiring empirical structural input. Searches are performed independently for each composition in the \ch{Li_{$y$}$M$O_{$x$}Cl_{6-2$x$}} series with $x = 0$, 0.5, 1, 1.5, 2, 2.5, and 3, and for the Group 5 metals Nb and Ta ($y = 1$) and the Group 4 metals Ti, Zr, and Hf ($y = 2$). Additional searches at $x = 0.67$, 0.75, 1.25, and 1.33 are performed for the Group 5 Nb and Ta series.

The search procedure follows the original SPINNER implementation~\cite{spinner} and uses the parameter set previously applied to Li-conducting oxide frameworks~\cite{topic}. Energy evaluations and structural optimizations during the search are performed with the SIMPLE-NN package~\cite{simplenn}, which implements a Behler--Parrinello-type neural-network potential~\cite{bpnn}. 8 radial and 18 angular symmetry functions are used for each element, with a cutoff radius of 6~\AA{}. The training sets are constructed using the melt--quench--annealing protocol with DFT simulations using the Vienna \textit{Ab initio} Simulation Package (VASP)~\cite{vasp} code. The MD parameters are selected following Hong \textit{et al}~\cite{csp_mqa}. We use projector augmented-wave pseudopotentials~\cite{paw} and the Perdew--Burke--Ernzerhof (PBE) exchange--correlation functional~\cite{pbe} with D3 dispersion corrections~\cite{D3,D3_2}. The plane-wave cutoff energies and $k$-point meshes are determined from convergence tests on premelted structures, ensuring that the total energies, forces, and stress tensors converge to within 20~meV/atom, 0.3~eV/\AA, and 10~kbar, respectively. The potential is trained before the structure searches and kept fixed throughout, and iterative retraining is not used.

In the structure search, candidate structures are generated by random seeding, permutation, and lattice mutation with relative frequencies of 70\%, 20\%, and 10\%, respectively. The population size is limited to 300 structures per generation, and each search is continued for 200 generations. In each generation, structures within 100~meV/atom of the lowest-energy structure in the preceding generation are selected for mutation operations. After completion of each search, candidates within 50~meV/atom of the lowest-energy structure are retained. The candidates are subsequently relaxed using SevenNet-Omni with D3 dispersion corrections. The ten lowest-energy structures at each composition are selected for subsequent analysis.

\subsection{Energy evaluation and thermodynamic stability}
\label{sec:methods_energy}
 
The structural relaxations of the SPINNER candidates and all MD simulations are performed with the pretrained universal MLIP, SevenNet-Omni, including Grimme D3 dispersion corrections. SevenNet-Omni is trained on several datasets computed with different DFT settings, and its output can be switched between them through a modality channel. Energies above the convex hull ($E_{\mathrm{hull}}$) are therefore evaluated with the channel trained on the Materials Project settings (denoted mpa)~\cite{CHGNet} and without dispersion corrections, so as to remain consistent with the reference energies of the Materials Project database~\cite{MaterialsProject} against which the hull is constructed.
 
\subsection{Molecular dynamics simulations}
\label{sec:methods_md}
 
Molecular dynamics simulations are performed in the canonical ($NVT$) ensemble using \texttt{LAMMPS}~\cite{lammps} with a timestep of 2~fs. One heavy metal atom is fixed in each simulation cell to remove spurious translational drift of the entire cell. This constraint is applied to the rigid metal--anion framework and does not affect the Li self-diffusion analysis. Supercells are constructed such that every lattice vector is at least 10~\AA{}. This criterion avoids interactions between a diffusing Li ion and its periodic image over the simulation timescale. The convergence of the calculated diffusivities with respect to cell size is confirmed in Fig.~S4.

The 600~K simulations are run for 100~ps. For the temperature-dependent simulations used to estimate room-temperature diffusivities, trajectories are extended to 200~ps at 550 and 500~K and to 400~ps at 450 and 400~K. This longer simulation time is used to compensate for the reduced frequency of Li hopping events and to obtain sufficient sampling for the diffusivity calculation. Three independent trajectories are performed for each structure and temperature.

\subsection{Li diffusivity from molecular dynamics}
\label{sec:methods_transport}

Li diffusivities are obtained from the mean-squared displacement (MSD) calculated from unwrapped MD trajectories. Before evaluating the MSD, all atomic positions are referenced to the instantaneous centroid of the host framework, which contains all non-Li atoms,
\begin{equation}
  \mathbf{r}'_i(t) = \mathbf{r}_i(t) - \frac{1}{N_{\mathrm{fw}}}
  \sum_{j \in \mathrm{fw}} \mathbf{r}_j(t),
  \label{eq:com}
\end{equation}
where $N_{\mathrm{fw}}$ is the number of framework atoms. Li displacements are thereby measured relative to the host lattice. This correction removes collective framework motion from the Li MSD. It is particularly important for the oxygen-rich layered structures ($x \ge 2$), in which rigid $M$--O--Cl slabs can undergo relative sliding at 600~K without loss of their internal framework connectivity. The same framework-referenced positions are used to evaluate the metal MSD for the framework-stability check.

The time-averaged Li MSD is calculated as
\begin{equation}
  \mathrm{MSD}(\tau) =
  \frac{1}{N_{\mathrm{Li}}N_{t_0}}
  \sum_{i=1}^{N_{\mathrm{Li}}}\sum_{t_0}
  \left|
  \mathbf{r}'_i(t_0+\tau)-\mathbf{r}'_i(t_0)
  \right|^2,
  \label{eq:msd}
\end{equation}
where the sum is taken over all Li ions and all available time origins $t_0$. The tracer diffusion coefficient is then obtained from the Einstein relation~\cite{einstein},
\begin{equation}
  D = \lim_{\tau \to \infty}
  \frac{\mathrm{MSD}(\tau)}{6\tau}.
  \label{eq:einstein}
\end{equation}
The slope is determined by linear regression over a lag-time window that begins when the MSD first exceeds 4.5~\AA$^2$, corresponding approximately to one Li-site displacement beyond the short-time vibrational regime. The fitting window ends at 70\% of the maximum lag time to retain sufficient time origins for statistical averaging~\cite{statistical}. Runs in which the Li MSD does not reach 4.5~\AA$^2$ are assigned $D = 0$. The factor of 6 in Eq.~\ref{eq:einstein} assumes isotropic three-dimensional diffusion. For a framework in which Li migrates along one- or two-dimensional pathways, the corresponding factor would be 2 or 4, so that $D$ would be larger by at most a factor of 3. This correction is small compared with the variation of $D$ across the oxygen-content series, which spans several orders of magnitude, and it does not affect the comparison between compositions or the identification of the fast-conduction window. All diffusivities are therefore reported with the three-dimensional factor, and the dimensionality of the actual pathways is analyzed separately through the diffusion tensor and $d_{\mathrm{eff}}$ (Sec.~\ref{sec:deff}). Each structure is simulated in three independent runs at every temperature. Reported diffusivities are the mean and standard deviation of the runs that retain an intact framework, defined by a framework-referenced metal MSD of no more than 5~\AA$^2$ over the trajectory. Diffusion tensors and $d_{\mathrm{eff}}$ (Sec.~\ref{sec:deff}) are calculated from the same framework-referenced Li displacements.

The ionic conductivity associated with the tracer diffusivity is estimated using the Nernst--Einstein relation~\cite{einstein},
\begin{equation}
  \sigma_{\mathrm{NE}} =
  \frac{\rho_{\mathrm{Li}} e^2 D}{k_{\mathrm{B}}T},
  \label{eq:ne}
\end{equation}
where $\rho_{\mathrm{Li}}$ is the Li number density, $e$ is the elementary charge, and $D$ is the Li tracer diffusion coefficient. Equation~\eqref{eq:ne} neglects correlation effects between mobile Li ions and is used here to relate the diffusivity screening criterion to an approximate ionic-conductivity scale.

For the 600~K screening, a structure is classified as a fast Li conductor when $D_{600\,\mathrm{K}} > 5\times10^{-6}$~cm$^2$/s. Assuming an activation energy of 0.3~eV, which is representative of the 0.2 to 0.4~eV range commonly reported for fast Li-ion conductors, this threshold corresponds to $D_{300\,\mathrm{K}} = 1.5\times10^{-8}$~cm$^2$/s. Using the Li number densities of the present compounds, $\rho_{\mathrm{Li}} = 0.6$ to $1.2\times10^{22}$~cm$^{-3}$ for $y = 1$ to 2, Eq.~\eqref{eq:ne} gives $\sigma_{300\,\mathrm{K}} \approx 0.6$ to 1~mS/cm. This room-temperature diffusivity benchmark is also comparable to NMR self-diffusivities reported for established solid electrolytes, including \ch{Li10GeP2S12}, \ch{Li6PS5Cl}, \ch{Li_{5.5}PS_{4.5}Cl_{1.5}}, and Ta-doped \ch{Li7La3Zr2O12}~\cite{LGPS,LPSC,LLZO}. These NMR measurements and the MD values are both tracer diffusion coefficients and can therefore be compared directly without a Haven-ratio correction.

\subsection{Framework dimensionality}
\label{sec:framework}

The dimensionality of the $M$--(O,Cl) framework is determined from a periodic structure graph. Metal atoms are treated as graph nodes, and edges correspond to the polyhedral links, together with their associated lattice-translation vectors. Li atoms and anions are not included as graph nodes. The resulting dimensionality therefore describes only the polyhedrally connected metal-centered framework and is independent of the Li sublattice.

The dimensionality of each connected component is determined using the algorithm of Larsen \textit{et al.}~\cite{framework} as implemented in pymatgen~\cite{pymatgen}. The method evaluates the rank of the independent lattice-translation vectors accumulated along cycles in the periodic graph. Components are classified as 0D for isolated units, 1D for chains, 2D for layers, and 3D for fully connected frameworks.

\subsection{Li free-energy landscape}
\label{sec:pmf}

The Li density $\rho(\mathbf{r})$ is accumulated on a real-space grid with a spacing of 0.2~\AA{} from framework-referenced Li positions [Eq.~\eqref{eq:com}] over the full MD trajectory. The positions are wrapped into the simulation cell before density accumulation. The potential of mean force is then calculated from
\begin{equation}
  F(\mathbf{r}) =
  -k_{\mathrm{B}}T
  \ln\left[
  \frac{\rho(\mathbf{r})}{\rho_{\max}}
  \right],
  \label{eq:pmf}
\end{equation}
where $\rho_{\max}$ is the maximum Li density sampled in the cell~\cite{pmf}. The PMF maps are displayed as projections onto a selected plane of the simulation cell. The color scale is truncated at 0.5~eV, and grid regions not visited by Li are assigned this maximum plotted value. The single-Li trajectories in Fig.~\ref{fig:mechanism}b show the PMF evaluated at the instantaneous position of one Li ion together with displacement from its initial position.

\subsection{Dimensionality of Li-ion transport}
\label{sec:deff}
The directional character of Li motion is quantified from the diffusion tensor, which is evaluated from the same framework-referenced, unwrapped trajectories and over the same fitting window as the scalar diffusivity,
\begin{equation}
  D_{ij} = \lim_{t\rightarrow\infty}
  \frac{\left\langle \Delta \mathbf{r}_i(t)\,\Delta \mathbf{r}_j(t)\right\rangle}{2t},
  \label{eq:Dtensor}
\end{equation}
where $\Delta \mathbf{r}_i$ is the $i$-th Cartesian component of the Li displacement and the average runs over all Li ions and time origins. The tensor is averaged over the independent runs of each structure. Diagonalization of $D_{ij}$ yields the principal diffusivities $D_1 \ge D_2 \ge D_3$, whose eigenvectors define the principal diffusion directions. The effective dimensionality of transport is then obtained from the normalized principal diffusivities $g_i = D_i / \sum_j D_j$ as
\begin{equation}
  d_{\mathrm{eff}} = \left(\sum_i g_i^{4}\right)^{-1/3}
  = \left[\frac{D_1^{4}+D_2^{4}+D_3^{4}}{(D_1+D_2+D_3)^{4}}\right]^{-1/3},
  \label{eq:deff}
\end{equation}
so that $d_{\mathrm{eff}} = 1$ corresponds to one-dimensional transport, 2 to isotropic two-dimensional transport, and 3 to isotropic three-dimensional transport.

Diagonalization is necessary because a migration channel that is not aligned with the simulation-cell axes gives similar values of $D_{xx}$, $D_{yy}$ and $D_{zz}$ even when the diffusion is strongly anisotropic (Fig.~S13a). The descriptor is validated on reference electrolytes with established transport anisotropy, where it recovers the one-dimensional channels of \ch{Li3YCl6} ($d_{\mathrm{eff}} = 1.0$), the quasi-two-dimensional vacancy layers of \ch{Li_{0.33}La_{0.56}TiO3} ($d_{\mathrm{eff}} = 1.7$) and the nearly isotropic network of \ch{Li7P3S11} ($d_{\mathrm{eff}} = 2.6$; Fig.~S13b)~\cite{dimensionality}.

\subsection{Electrochemical stability window}
\label{sec:methods_esw}

Electrochemical stability windows are evaluated from Li grand potential phase
diagrams~\cite{Richards2016}. The energies are those used for the convex hull, and the competing phases are all entries of the \ch{Li}--$M$--\ch{O}--\ch{Cl} chemical space taken from the Materials Project~\cite{MaterialsProject}. The grand potential of a phase,
$\Phi = E - \mu_{\mathrm{Li}} N_{\mathrm{Li}}$, is evaluated at Li chemical potentials referenced to metallic Li, which correspond to applied voltages of $V = -\mu_{\mathrm{Li}}/e$, and the grand potential hull is constructed between 1 and 5~V in steps of 0.05~V. The window is the voltage range over which the composition remains on the hull, and its limits are the onset potentials at which decomposition into the most favorable products becomes thermodynamically favorable. Since the predicted structures lie slightly above the convex hull, their energies are referenced to the hull beforehand, so that the window describes the composition itself.

\section*{Acknowledgements}
This research was supported by the Nano \& Material Technology Development Program through the National Research Foundation of Korea (NRF) funded by Ministry of Science and ICT (RS-2024-00407995 and RS-2023-00247245). The computations were carried out at Korea Institute of Science and Technology Information (KISTI) National Supercomputing Center (KSC-2025-CRE-0110).

\section*{Declaration of competing interest}
The authors declare that they have no known competing financial interests or personal
relationships that could have appeared to influence the work reported in this paper.

\section*{Supplementary material}
Supplementary material associated with this article can be found in the online version.

\section*{Data availability}

Data will be made available on request.

\bibliographystyle{model1-num-names}
\bibliography{refs}

\end{document}